\documentclass[runningheads]{llncs}
\usepackage{graphicx}
\usepackage{changepage}% http://ctan.org/pkg/changepage
\usepackage{lipsum}% http://ctan.org/pkg/lipsum
\usepackage{xcolor}
\usepackage[breaklinks, colorlinks, allcolors=blue]{hyperref}
\usepackage{floatrow}
\usepackage{tabularx}
\usepackage{amssymb}
\usepackage{amsmath}
\usepackage{mathrsfs}
\usepackage[normalem]{ulem}  
\usepackage{hyperref}
\usepackage{url}
\usepackage{ctable} % for \specialrule command
\usepackage{tabularx}
\usepackage{adjustbox}
\usepackage{booktabs}
\usepackage{array}

\usepackage[nocompress]{cite} % order citations

\usepackage[ampersand]{easylist}
\newcolumntype{?}{!{\vrule width 1pt}}

\newcommand{\rc}[1]{{\textcolor{black}{{#1}}}}

\makeatletter
\def\thickhline{%
  \noalign{\ifnum0=`}\fi\hrule \@height \thickarrayrulewidth \futurelet
  \reserved@a\@xthickhline}
\def\@xthickhline{\ifx\reserved@a\thickhline
              \vskip\doublerulesep
              \vskip-\thickarrayrulewidth
             \fi
      \ifnum0=`{\fi}}
\makeatother

\newlength{\thickarrayrulewidth}
\usepackage{bigfoot}
\DeclareNewFootnote[para]{default}

\begin{document}

\title{Figure and Figure Caption Extraction for Mixed Raster and Vector PDFs: Digitization of Astronomical Literature with OCR Features}% \thanks{This work supported by NASA Astrophysics Data Analysis Program Grant (20-ADAP20-0225). }}
\titlerunning{Figure and Figure Caption Extraction}
\author{J. P. Naiman\inst{1}\orcidID{0000-0002-9397-6189} \and
Peter K. G. Williams\inst{2}\orcidID{0000-0003-3734-3587} \and
Alyssa Goodman\inst{2}\orcidID{0000-0003-1312-0477}}
\authorrunning{J.P. Naiman et al.}
\institute{School of Information Sciences, University of Illinois, Urbana-Champaign, 61820, USA
\email{jnaiman@illinois.edu}\\
%\url{http://www.springer.com/gp/computer-science/lncs} 
\and
Harvard-Smithsonian Center for Astrophysics, Cambridge, 02138, USA\\
\email{\{pwilliams,agoodman\}@cfa.harvard.edu}}
\maketitle              % typeset the header of the contribution
\begin{abstract}
    Scientific articles published prior to the ``age of digitization" in the late 1990s contain figures which are ``trapped" within their scanned pages.  
    While progress to extract figures and their captions has been made, there is currently no robust method for this process. 
    %Here, w
    We present a YOLO-based method for use on scanned pages, post-Optical Character Recognition (OCR), which uses both grayscale and OCR-features.  When applied to the astrophysics literature holdings of the Astrophysics Data System (ADS), we find F1 scores of 90.9\% (92.2\%) for figures (figure captions) with the intersection-over-union (IOU) cut-off of 0.9 which is a significant improvement over other state-of-the-art methods. 

\end{abstract}

\section{Introduction}

With the rise of larger datasets and the ever increasing rate of scientific publication, scientists require the use of automated methods to parse these growing data products, including the academic literature itself.  
In addition to being a vital component of open science \cite{sandy2017,sohmen2018figures}, easily accessed and well curated data products are strongly encouraged as part of the submission process to most major scientific journals \cite{mayernik2017}.
However, data products in the form of figures, tables and formulas which are stored in the academic literature, especially from the ``pre-digital" era, published prior to $\sim$1997, are not accessible for curation unless methods are developed to extract this important information.

The extraction of different layout elements of articles is an important component of scientific data curation, with the accuracy of extraction of the elements such as tables, figures and their captions increasing significantly over the past several years \cite{icdar2017,podreview3,podreview1,lehenmeier2020layout}.
A large field of study within document layout analysis is the ``mining" of PDFs as newer PDFs are generally in ``vector" format -- the document is rendered from a set of instructions instead of pixel-by-pixel as in a raster format, and, in theory, the set of instructions can be parsed to determine the locations of figures, captions and tables \cite{klampfl2013unsupervised,bai2006automatic,choudhury2013figure}.

However, this parsing is non-trivial and many methods have been developed to complete this task. If the PDF's vector format is well structured, then text and images can be extracted by parsing this known PDF format.  Several packages exist which output text and/or images from such PDF files \cite{GROBID}.
Historically, some of the most popular include heuristic methods where blocks of text are classified as figure or table captions based on keywords (like ``Fig." or ``Figure") \cite{Choudhury2013,pdffigures2}.  

Deep learning methods have become popular recently for vector and raster documents \cite{surveydeeplearning,deepfigures}, including those that use methods of semantic segmentation \cite{yang2017layout} and object detection \cite{saha2019}. While these methods are vital to the extraction of data products from recent academic literature, pre-digital literature is often included in digital platforms with older articles scanned at varying resolutions and deep learning methods developed with newer article training sets often perform poorly on this pre-digital literature \cite{scanbank}.
Additionally, layouts, fonts, and article styles are typically different for historical documents when compared to ``born-digital" scientific literature \cite{scanbank}.
In these cases, text extraction must be performed with optical character recognition (OCR), and figures and tables are extracted from the raw OCR results.
When applied to raster-PDF's with text generated from OCR, deep learning document layout analysis methods trained with newer or vector-based PDFs are often not as robust \cite{yang2017layout,scanbank}.
While progress has been made in augmenting these methods for OCR'd pages, especially for electronic theses and dissertations (ETDs) \cite{scanbank}, much work can still be done to extract layout elements from these older, raster-based documents.

In what follows, we outline a new methodology for extracting figures and figure captions from a data set that includes both vector and raster based PDF's from the pre-digital %(prior to 1997) 
scientific literature holdings of the Astrophysics Data System (ADS)\footnote{\url{https://ui.adsabs.harvard.edu/}}.  
Our model applies deep learning object detection methods in combination with heuristic techniques to scans of article pages as well as the text features generated from processing scans through the \textsf{Tesseract} OCR engine \cite{tesseract} and combines the results from mining any vector based PDF's for their captions with \textsf{pdffigures2} \cite{pdffigures2} in a post-processing step.

While the focus of our model is the digitization of astronomical literature -- one of the original ``big data" sciences \cite{astrobigdata1,astrobigdata2} -- because our method relies heavily on features generated with OCR, our methodology is extendable to other scientific literature. 
Additionally, the design of our pipeline is heavily motivated by both the data (astronomical literature) and the expected users of our model (scientists and digital librarians).  Thus, we rely on open-source software (e.g. \textsf{Tesseract}) and programming languages used by both communities (e.g. \textsf{Python}). All code is available on GitHub\footnote{\url{https://github.com/ReadingTimeMachine/figure_and_caption_extraction}}.

\section{The Data} \label{section:data}

The dataset used in this work is a subset of the English-language literature archived on ADS including pre-digital articles published between $\approx$1850-1997.  Chosen for this work are scanned article pages that are thought to contain images of the sky derived from heuristic determinations based on color distributions \cite{adsass2012}.

We begin by defining the classes of figure and figure caption as there is often disagreement in the literature and occasionally between annotators of the same dataset \cite{icdar2017,younas2019}. 
In this work, we define a figure as the collection of one or more panels on a single page which would be referred to as a single figure in a scientific article (i.e. ``Figure 3").  This is different than other works which often treat each ``sub-figure" as a separate figure \cite{younas2019}.
Additionally, figures are defined to include all axis labels and titles as well as the area above their associated captions (or are extended on the side facing the caption if the caption is not below the figure).  Finally, if a figure caption extends horizontally further than its associated figure, the figure is extended horizontally to the edges of the figure caption (see magenta lines in \autoref{fig:postprocessing}).  These definitions retain the uniformity of other definitions (e.g.\ \cite{publaynet}) while defining figure and caption regions by non-overlapping boxes.

Using these definitions, we hand-annotate images to create ground-truth boxes using the \textsf{MakeSense.ai}\footnote{\url{makesense.ai}} annotation tool for a total of 5515 pages which contain 5010 figures and 4925 figure captions.
 
There are 553 pages without figures or captions.  Details about the annotation processing is the subject of an upcoming paper \cite{naimanprep}.  

Each annotated page is also processed with \textsf{Tesseract}'s OCR engine and outputs are stored in the \textsf{hOCR} format.  Additionally, we process the PDF version of the article associated with each scanned page through the PDF-mining software \textsf{pdffigures2} \cite{pdffigures2} and store any found figures and figure captions for combination with our model's results in a post-processing step (see \rc{Step 3 in} \autoref{section:postprocessing}).

In what follows, we develop a pipeline that relies heavily on features derived from the OCR results of raster-PDF's to make use of the preponderance of these types of PDFs in our dataset.

\vspace{-1mm}
\section{Model Pipeline development}

\begin{figure}%[h]
\centering
\includegraphics[width=0.99\textwidth]{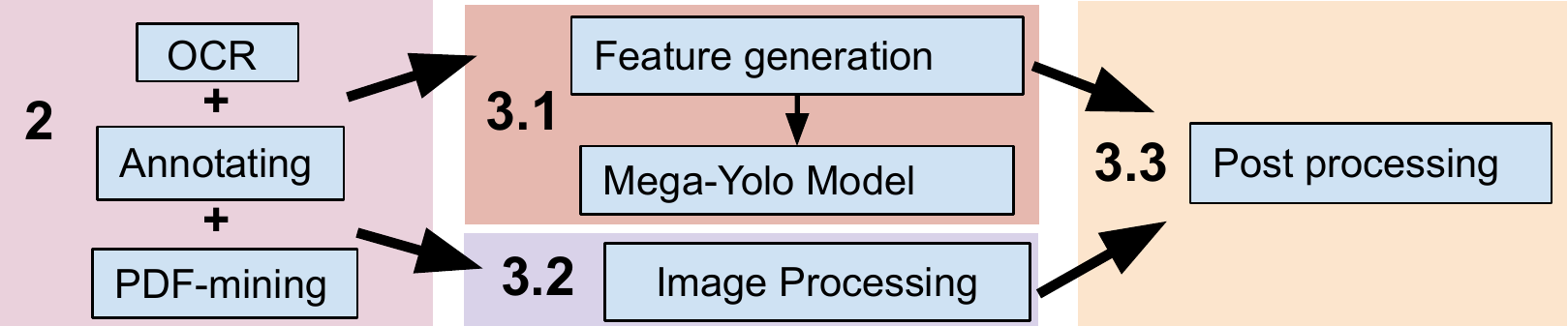}
\vspace*{-2mm} % kosher?
\caption{Our overall pipeline is shown as four main steps. Data processing and annotating is discussed in \autoref{section:data}, feature selection and deep learning model description is housed in \autoref{section:deeplearningmodel}, image processing techniques for heuristic figure-finding in \autoref{section:imageprocessing}, and post-processing techniques are discussed in  \autoref{section:postprocessing}. }
\label{fig:pipeline}
\end{figure}

Our final goal for this dataset is the hosting of figure-caption pairs on the Astronomy Image Explorer (AIE) database\footnote{\url{http://www.astroexplorer.org/}}.  Currently, a subset of the born-digital ADS holdings (articles housed in under the American Astronomical Society Journals (AAS)) automatically populate AIE with their figure-caption pairs.
The \textsf{Python}-based pipeline developed working toward this goal is shown in \autoref{fig:pipeline}.

\subsection{Deep learning model and Feature Selection} \label{section:deeplearningmodel}

Typical modern methods rely on deep learning techniques to detect layout elements on pages \cite{surveydeeplearning}, often in combination with heuristics \cite{deepfigures}.
Methods span the range of object detection using models like YOLO \cite{yolo1,deepfigures,scanbank} to, more recently, Faster R-CNN \cite{fasterrcnn,fasterrcnnDeepDesert,vo2018,younas2019,fasterrcnntables} and Mask-CNN \cite{maskrcnn,maskrcnnDocbank,maskrcnncdec}. Additionally, several pixel-by-pixel segmentation models have been proposed using semantic segmentation \cite{yang2017layout} and fully convolutional networks \cite{fcnn,fcnncharts}, including ``fully convolutional instance segmentation" \cite{fcis1,fcis2,fcis3}.

Often these models employ a variety of features derived from article pages as inputs along side or in place of the unprocessed page.
Some of the more popular recent methods leverage image processing and computer vision techniques \cite{younas2019} including connected component analysis \cite{connectedComp,conncomp2,conncomp3,conncomp4} and the distance transform \cite{fasterrcnntables} or some combinations thereof \cite{fifo}. %see these in more detail too!

While the aim of many methods is to detect page objects before any OCR process, here we implement methods that can be applied after.
In what follows, we use a \textsf{Tensorflow} implementation of YOLO-v5\footnote{\url{https://github.com/jahongir7174/YOLOv5-tf} \\ \url{https://github.com/jmpap/YOLOV2-Tensorflow-2.0}} \cite{yolov1,yolov5} and focus our efforts on feature exploration by utilizing a set of features derived from the OCR outputs themselves with the goal to choose the smallest number of the ``best" features on which to build our model.  
In addition to the raw grayscale page, there are several possible features derived primarily from the \textsf{hOCR} outputs of \textsf{Tesseract}.
To minimize storage, each feature is scaled as an unsigned-integer, 8-bit ``color" channel in a \rc{512x512 (pixels),} multi-layer image which is fed into a ``mega-YOLO" model capable of processing more than three color channels.

%\clearpage % kosher?
\noindent Features explored which are output from \rc{\textsf{Tesseract}} in  \textsf{hOCR} format  include:
\vspace{-2mm} % kosher?
\begin{itemize}
    \item {\it{fontsize (fs)}}: the fontsize for each word bounding box is normalized by subtraction of the median page fontsize and division by the standard deviation.  Bounding boxes with fonts outside five standard deviations are ignored.
    \item {\it carea ($c_{\rm b}$)}: the ``content area" from automatic page segmentation includes large blocks of text and sometimes encapsulates figures into separate ``content areas", but not consistently.
    \item {\it paragraphs ($p_{\rm b}$)}: automatically segmented groups of words as likely paragraphs. Often overlaps with ``carea".
    \item {\it ascenders (asc)}: from typography definitions -- the amount of letters in the word that are above the letter ``caps" (e.g. the top of the letter ``h").  Ascenders are normalized by subtracting the median value for each page.
    \item {\it decenders (dec)}: a typographical element -- the amount of letters in the word that are below the letter ``bottoms" (e.g. the bottom curl of the letter ``g").  Decenders are normalized by subtracting the median value for each page.
    \item {\it word confidences (wc)}: the percent confidence of each word 
    \item {\it word rotation (t$_{\rm ang}$)}: rotation of word in steps of 0$^\circ$, 180$^\circ$, and 270$^\circ$.
\end{itemize}

\noindent Other features derived from the page scan and OCR-generated text are:
\vspace{-2mm} % kosher?
\begin{itemize}
    \item {\it grayscale (gs)}: the image is collapsed into grayscale using the page's luminance. The majority of images are already in grayscale and those few that are in color are transformed to grayscale.
    \item {\it fraction of letters in a word (\%l)}: the percentage of characters in a word that are letters (scaled 125-255 in order to preserve a ``true zero" for spaces in the scanned page that contain no words).
    \item {\it fraction of numbers in a word (\%n)}: the percentage of numbers in a word that are letters (scaled 125-255).
    \item {\it punctuation (p)}: punctuation marks are tagged as 250, non-punctuation characters are tagged as 125 (saving 0 for empty, non-word space). 
    \item {\it spaCy POS (SP)}: spaCy's \cite{spacy2} 19 entries for ``part of speech" (noun, verb, etc) in the English language
    \item {\it spaCy TAG (ST)}: more detailed part of speech tag with 57 values
    \item {\it spaCy DEP (SD)}: the 51 ``syntactic dependency" tags which specify how different words are related to each other
\end{itemize}

\begin{figure}%[h]
\centering
\includegraphics[width=1.0\textwidth]{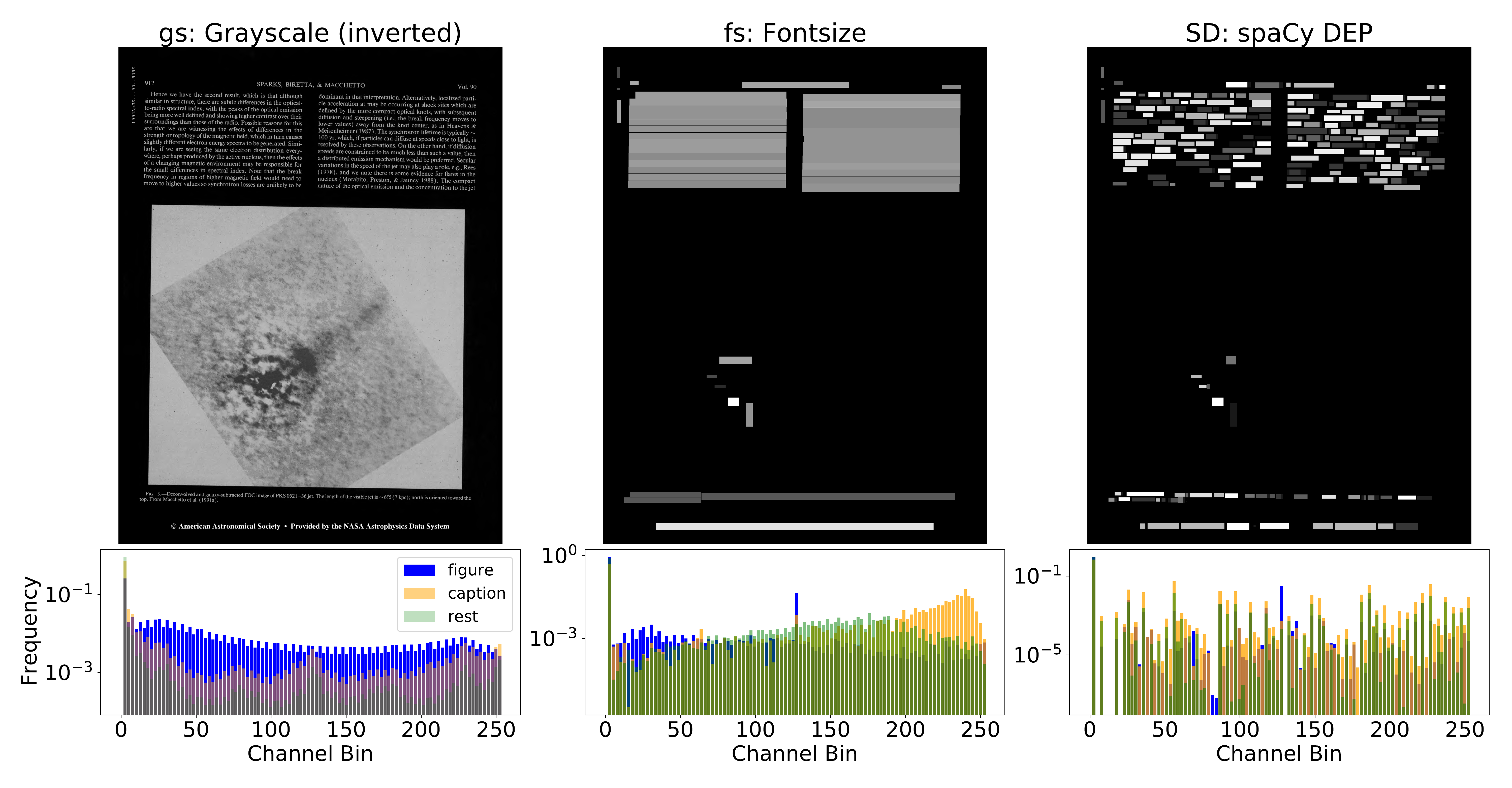}
\vspace*{-7mm} % kosher?
\caption{Examples of selected features for a single page (top row) and the distribution across all pages in the annotated dataset (bottom row).  All features have been rescaled into 8-bit color bins (see \autoref{section:deeplearningmodel}). Here, ``rest" refers to rest of the page that does not include figures or captions.  Several features show clear differences in distributions (e.g. grayscale and fontsize) while others do not (e.g. spaCy DEP). }
\label{fig:features}
\end{figure}

\autoref{fig:features} shows an example of a selection of these features (grayscale ({\it gs}), fontsize ({\it fs}), and spaCy DEP ({\it SD})) for a single page and their distributions across all pages in our annotated dataset.  
The example features of grayscale and fontsize show differences in distributions in the three categories of figure, figure caption and the ``rest" of the page -- grayscale distributions are more uniform inside figures (left plot of \autoref{fig:features}) and figure captions show a peak in the fontsize distributions toward higher values when compared to the fontsize distributions of figures (middle plot of \autoref{fig:features}). Trends in other features are harder to determine, as illustrated in the bottom right panel of \autoref{fig:features} which shows a less clear distinction between figures, figure captions, and the rest of the page for the feature of spaCy DEP.

In \autoref{section:results} we discuss our best model which includes (grayscale,  ascenders, decenders, word confidences, fraction of numbers in a word, fraction of letters in a word, punctuation, word rotation and spaCy POS) as the set of input features.

\vspace{-3mm} % kosher?
\subsection{Image Processing} \label{section:imageprocessing}
%\vspace{-2mm} % kosher?

In addition to the deep learning model described in \autoref{section:deeplearningmodel}, we use image processing techniques to heuristically find potential figure boxes to combine with those found with the mega-YOLO model.
The locations of the OCR'd words (see \autoref{section:data}) are used to mask out text and the modified pages are processed through a basic shape finder built with OpenCV, tuned to look for rectangles (four corners and sides comprised of approximately parallel lines).
This ``rectangle finder" is applied to several filtered versions of the page (histogram of oriented gradients, dilation, and color-reversal, and various levels of thresholding). The list of rectangles is culled with K-Means clustering on the locations of square corners, checking for artifact-rectangles which are small in size, and rectangles that are likely colorbars and not figures due to their aspect ratio. 

OCR'ing a page and shape-finding with OpenCV takes approximately 20-25 seconds per page (tested on six cores of an Apple M1 Max with 64 Gb of RAM).

\vspace{-1mm}
\subsection{Post-Processing Pipeline} \label{section:postprocessing}

After features are selected and the model is trained we modify the final found boxes by merging them with OCR word and paragraph boxes and any heuristically found captions and figures \rc{at the fractional-pixel level (results are rounded to nearest pixel for intersection-over-union (IOU) calculations to match precision of ground truth boxes)}.
Post-processing is a common practice in document layout analysis \cite{fifo,yi2017}, however it often differs between implementations and is occasionally not incorporated into a final pipeline \cite{wu2019detectron2}.
\begin{figure}%[h]
\centering
\includegraphics[width=0.95\textwidth]{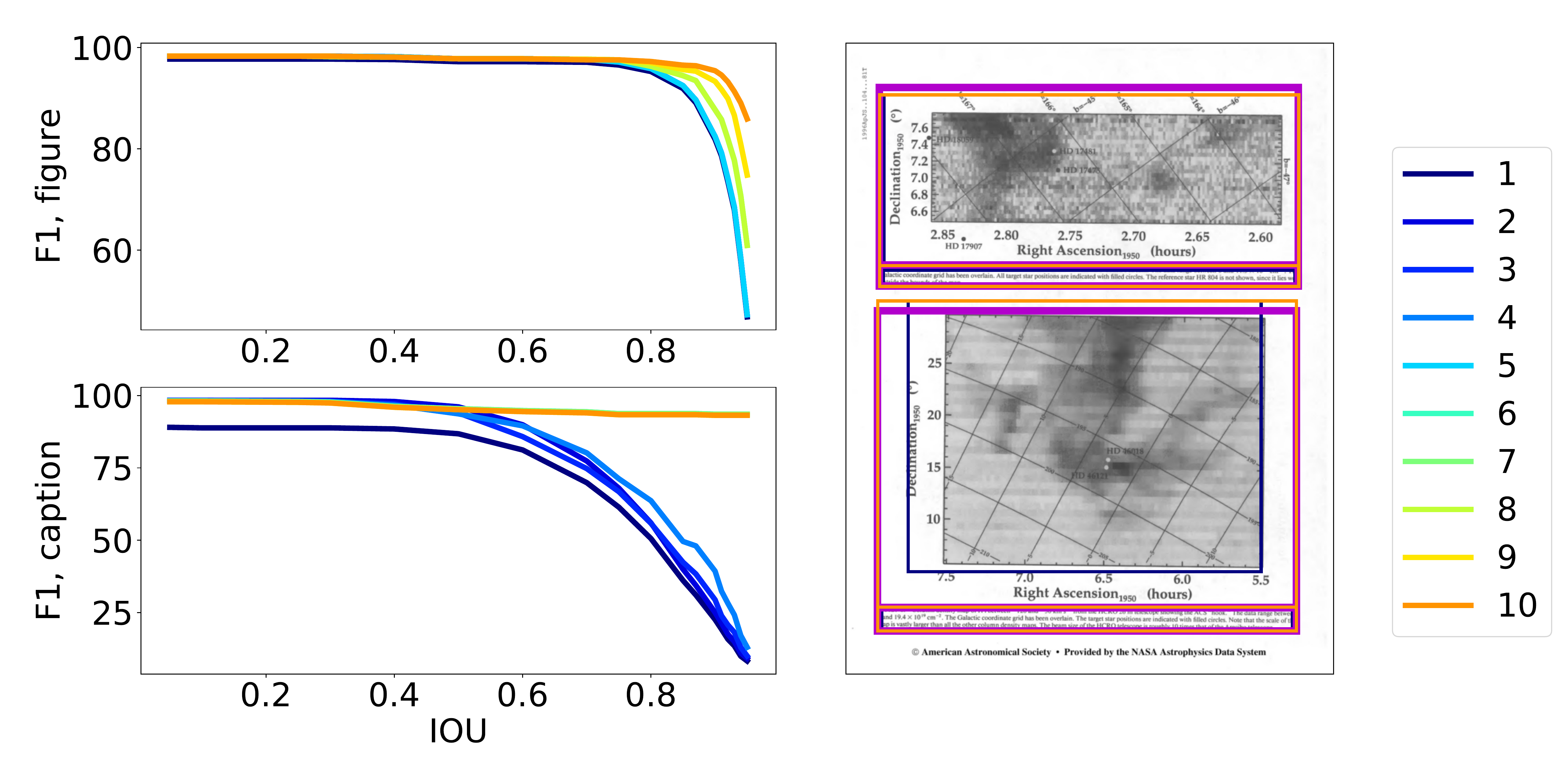}
\vspace{-5mm}
\caption{Effects of post-processing steps on \rc{F1 (left plots)} for Model 12 (m12 in \autoref{tab:features} and \autoref{tab:featuresSelection}).  Post-processing drives changes in the metrics at larger IOU's -- IOU$\gtrsim$0.8 and IOU$\gtrsim$0.6 for figures and captions, respectively.  \rc{Changes are depicted for a single page (right plot) showing initial found boxes (Step 1, dark blue) and final (Step 10, orange) in comparison to true boxes (thick magenta).}}
\label{fig:postprocessing}
\end{figure}
\autoref{fig:postprocessing} depicts how \rc{found boxes and} %precision, recall, and 
F1 score changes with each post-processing step in our pipeline when we compare ground-truth (true) boxes to model-found boxes at various post-processing steps:
\vspace{-2mm}
\begin{itemize}
    \item Step 1: ``raw" found boxes are those culled with non-maximum suppression  
    \item Step 2: if two found boxes overlap with an IOU~$\ge$~0.25 the box with the lowest score is removed, decreasing false positives (FP)
    \item Step 3: \textsf{pdffigures2}-found figure caption boxes replace those found with the deep learning model when they overlap which increases caption true positive (TP) rate and decreases FP and false negative (FN) at large IOU thresholds.
    \item Step 4: Caption boxes are found heuristically by first applying a gaussian blur filter on an image of drawn OCR text boxes. Contours of this image that overlap with text boxes that match with a fuzzy-search of words such as ``Fig.", ``Figure" and ``Plate" are labeled as heuristically-found.  If a heuristically-found caption box overlaps with a mega-YOLO-found box, we take the top of the heuristic box (which tends to be more accurate) and the minimum (maximum) of the left (right, bottom) of the two boxes.  This results in an overall increase in TP while FN and FP drop.  
    \item Step 5: found captions are expanded by their overlap with OCR word and paragraph boxes, allowing for multiple ``grow" iterations in the horizontal direction.  Found boxes are only expanded by paragraph and word boxes if the centers of paragraph and word boxes overlap with the found box.
    \item Step 6: if found figure boxes overlap with rectangles that are found through image processing (as described in \autoref{section:imageprocessing}), the found box is expanded to include the image processing rectangle.  This increases the TP rate at larger IOU thresholds for figures.
    \item Step 7: any found captions that have areas larger than 75\% of the page area are discarded leading to a slight drop in FP for captions.
    \item Step 8: Captions are paired to figures by minimizing the distance between caption center and bottom of a specific figure.  Rotational information from the page and overall rotation of the OCR words is used to determine the ``bottom" of the figure. Any captions without an associated figure on a page are dropped, leading to a drop in FP.
    % \item Step 9: both true and found figure boxes are extended down to the tops of their associated captions increasing TP for figures and captions at high IOU thresholds.
    % \item Step 10: if a figure caption extends horizontally further than its associated figure, the figure is extended horizontally to the edges of the figure caption for both true and found boxes.  This leads to an increase in TP rates for figures at high IOU thresholds.
    \item Step 9: found figure boxes are extended down to the tops of their associated captions increasing TP for figures and captions at high IOU thresholds.
    \item Step 10: if a figure caption extends horizontally further than its associated figure, the figure is extended horizontally to the edges of the figure caption.  This leads to an increase in TP rates for figures at high IOU thresholds.
    
    \end{itemize}

Steps 9 and 10 are similar to the steps described for annotated boxes in \autoref{section:data}. The effects to the metrics shown in Figure \ref{fig:postprocessing} are modest and predominately affect the results at high IOU thresholds (IOU$\gtrsim$0.9) for figures.

%\vspace{-5mm}
\subsection{Feature Selection Ablation Experiments } \label{section:experiments}
To determine the set of features which produce the most accurate model while minimizing the feature memory footprint, we conduct a series of ablation experiments, summarized in \autoref{tab:features} and \autoref{tab:featuresSelection}.
In all feature selection runs we use 75\% of our data in the training set, 15\% in validation, and 10\% in the test dataset.  Results in \autoref{tab:featuresSelection} are shown for this feature selection test dataset.

\begin{table}[!htbp]
\small
\begin{center}
\begin{tabular}{|c|l|}
    \hline
    model & Description \\
    \thickhline 
    m1 & gs  \\
    \hline
    m2 & gs + fs \\
    \hline
    m3 & gs + fs + asc + dec \\
    \hline
    m4 & gs + fs + asc + dec + wc \\
    \hline
    m5 & gs + fs + asc + dec + wc + \%n  + \%l + p\\
    \hline
    m6 & gs + fs + asc + dec + wc + \%n  + \%l + p + $\rm{t_{ang}}$  \\
    \hline
    m7 & gs + fs + asc + dec + wc + \%n  + \%l + p + $\rm{t_{ang}}$ + SP  \\
    \hline
    m8 & gs + fs + asc + dec + wc + \%n  + \%l + p + $\rm{t_{ang}}$ + SP + ST + SD   \\
    \hline
    m9 & gs + fs + asc + dec + wc + \%n  + \%l + p + $\rm{t_{ang}}$ + SP + ST + SD + $\rm{p_{b}}$  \\
    \hline
    m10 & gs + fs + asc + dec + wc + \%n  + \%l + p + $\rm{t_{ang}}$ + SP + ST  + SD + $\rm{p_{b}}$ + $\rm{c_{b}}$  \\
    \hline
    \thickhline
    m11 & gs + fs + wc +  \%n  + \%l  + p + $\rm{t_{ang}}$ + SP  \\
    \hline
    \bf m12 & \bf gs + asc + dec + wc + \%n  + \%l  + p + $\rm{t_{ang}}$ + SP  \\    %\hline 
    \hline
    m13 & gs + asc + dec + wc + \%n  + p + $\rm{t_{ang}}$ + SP  \\    %\hline 
    \hline
    m14 & gs + asc + dec + wc + \%n  + \%l  + $\rm{t_{ang}}$ + SP  \\    
    \hline
    m15 & gs + asc + dec + wc + \%n  + \%l  + p + $\rm{t_{ang}}$ \\    
    %model Y & include HOG, etc from image processing as other layers of features \\

    \hline
\end{tabular}
\end{center}
\vspace{-5mm}
\caption{ Ablation experiments with the features discussed in \autoref{section:deeplearningmodel}. All models include post-processing (\autoref{section:postprocessing}).  Our ``best" model, as determined by metrics in \autoref{tab:featuresSelection} and the discussion of \autoref{section:experiments}, is Model 12 (m12) highlighted in bold.} 
\label{tab:features}
\end{table} 

As it is computationally prohibitive to test all combination of all fourteen different features, we first adopt the strategy of including sets of one or two groups of features at a time until we have a model containing all fourteen features, as shown above the thick horizontal line in \autoref{tab:features}.
From these ten models, we select the most accurate, defined here as having a high F1 score for both figures and their captions, while maintaining a low false positive score (FP).  Model 7 is the ``best" model out of these first ten models in \autoref{tab:featuresSelection}. 
We then subtract one or two features from this model in combinations shown below the thick horizontal line in \autoref{tab:features}.  Using the same selection criteria leads us to choose Model 12 as our overall ``best" model which includes the features of (grayscale,  ascenders, decenders, word confidences, fraction of numbers in a word, fraction of letters in a word, punctuation, word rotation and spaCy POS) as highlighted in \autoref{tab:featuresSelection}.

\rc{  
The implemented optimizer is Adam with a $\beta_1=0.937$ $\beta_2=0.999$. 
Learning rate is scheduled using a cosine scheduler which depends on initial learning rate, number of epochs and batch size. Practically, when applied to our model this results in a linear increase in learning rate by a factor of $\sim$1.6 in the initial epoch (flat after). Our optimal initial learning rate of 0.004 was chosen from a small set of learning rates (0.008, 0.004, 0.0004, 0.0002). All experiments are run for 150 epochs and converge within this time (tracked by validation losses).} No data augmentation is applied. Training is performed on a Tesla V100-SXM2 GPU with an average time of $\sim$6.5~minutes per epoch.

% features table
\begin{table}[!htbp]
%\vspace{-10mm} % kosher?
\footnotesize
\begin{center}
%\begin{tabular}{|c?cc?cccc?cc|}
\begin{tabular}{|c?cc|cc|cc?cc|cc|cc|}
    \hline
    & \multicolumn{2}{c|}{TP}  & \multicolumn{2}{c|}{FP} & \multicolumn{2}{c?}{FN} & \multicolumn{2}{c|}{Prec} & \multicolumn{2}{c|}{Rec} & \multicolumn{2}{c|}{F1} \\
     %& figure & caption & figure & caption & figure & caption & figure & caption & figure & caption & figure & caption \\
     & fig & cap & fig & cap & fig & cap & fig & cap & fig & cap & fig & cap \\

    \hline

    m1 & 90.3\unskip & 88.3\unskip &  11.3\unskip & 10.0\unskip & 1.8\unskip & 4.3\unskip & 88.9\unskip & 89.8\unskip & 98.0\unskip & 95.4\unskip & 93.3\unskip & 92.5\unskip \\

    m2 & 89.5\unskip & 86.9\unskip &  10.9\unskip & 8.6\unskip & 2.6\unskip & 6.4\unskip & 89.2\unskip & 91.0\unskip & 97.2\unskip & 93.2\unskip & 93.0\unskip & 92.1\unskip \\

    m3 & 85.9\unskip & 86.9\unskip &  17.9\unskip & 12.5\unskip & 1.8\unskip & 5.1\unskip & 82.8\unskip & 87.4\unskip & 97.9\unskip & 94.4\unskip & 89.7\unskip & 90.8\unskip \\

    m4 & 90.5\unskip & 88.7\unskip &  10.1\unskip & 8.8\unskip & 2.0\unskip & 3.7\unskip & 90.0\unskip & 91.0\unskip & 97.8\unskip & 96.0\unskip & 93.7\unskip & 93.4\unskip \\

    m5 & 84.5\unskip & 87.3\unskip &  15.7\unskip & 7.2\unskip & 2.2\unskip & 7.0\unskip & 84.3\unskip & 92.4\unskip & 97.4\unskip & 92.6\unskip & 90.4\unskip & 92.5\unskip \\

    m6 & 89.5\unskip & 89.1\unskip &  11.9\unskip & 9.0\unskip & 2.0\unskip & 4.3\unskip & 88.3\unskip & 90.8\unskip & 97.8\unskip & 95.4\unskip & 92.8\unskip & 93.0\unskip \\

    m7 & 92.8\unskip & 88.1\unskip &  8.0\unskip & 9.0\unskip & 1.4\unskip & 4.1\unskip & 92.0\unskip & 90.7\unskip & 98.5\unskip & 95.6\unskip & 95.1\unskip & 93.1\unskip \\

    m8 & 90.5\unskip & 90.0\unskip &  9.1\unskip & 7.2\unskip & 2.0\unskip & 3.9\unskip & 90.9\unskip & 92.6\unskip & 97.8\unskip & 95.9\unskip & 94.2\unskip & 94.2\unskip \\
    
    m9 & 84.3\unskip & 84.2\unskip &  13.5\unskip & 7.4\unskip & 4.0\unskip & 9.4\unskip & 86.2\unskip & 91.9\unskip & 95.4\unskip & 89.9\unskip & 90.6\unskip & 90.9\unskip \\
    
    m10 & 88.7\unskip & 87.5\unskip &  11.5\unskip & 9.6\unskip & 1.8\unskip & 4.3\unskip & 88.6\unskip & 90.1\unskip & 98.0\unskip & 95.3\unskip & 93.0\unskip & 92.6\unskip \\
    m11 & 90.5\unskip & 92.4\unskip &  10.5\unskip & 6.8\unskip & 0.8\unskip & 1.8\unskip & 89.6\unskip & 93.2\unskip & 99.1\unskip & 98.0\unskip & 94.1\unskip & 95.6\unskip \\
    \bf m12 & \bf 92.2\unskip & \bf 89.1\unskip &  \bf 6.4\unskip & \bf 6.6\unskip & \bf 2.4\unskip & \bf 4.9\unskip & \bf 93.5\unskip & \bf 93.1\unskip & \bf 97.4\unskip & \bf 94.8\unskip & \bf 95.4\unskip & \bf 94.0\unskip \\
    m13 & 92.8\unskip & 88.7\unskip &  7.8\unskip & 8.4\unskip & 2.0\unskip & 4.3\unskip & 92.2\unskip & 91.4\unskip & 97.9\unskip & 95.4\unskip & 95.0\unskip & 93.3\unskip \\

    m14 & 87.3\unskip & 88.7\unskip &  15.9\unskip & 8.4\unskip & 1.2\unskip & 6.4\unskip & 84.6\unskip & 91.4\unskip & 98.6\unskip & 93.3\unskip & 91.1\unskip & 92.3\unskip \\

    m15 & 89.9\unskip & 89.5\unskip &  8.7\unskip & 5.9\unskip & 2.4\unskip & 5.1\unskip & 91.2\unskip & 93.8\unskip & 97.4\unskip & 94.6\unskip & 94.2\unskip & 94.2\unskip \\

    \hline
\end{tabular}
\end{center}
\vspace{-5mm} % kosher?
\caption{Metrics for models described in \autoref{tab:features}. There are {\protect 497\unskip} figures (fig) and {\protect 488\unskip} figure captions (cap) used in the feature selection test dataset. IOU is 0.9 for both figures and captions.   
TP, FP and FN are shown as percentages of the total instances.% in each category.
}
%\vspace{-10mm}
\label{tab:featuresSelection}
%\vspace{-10mm}
\end{table}

%\begin{multicols}{2}

\vspace{-10mm} % kosher?
\section{Results} \label{section:results}

To quantify the results of our ``best" model (Model 12) on un-seen data we annotate an additional $\approx$600 pages as a ``final test dataset" of $\approx$500 figure and figure caption ground-truths (490 and 487, respectively). We find F1 scores of 90.9\% (92.2\%) for figure (caption) detections at an intersection-over-union cut off (IOU) of 0.9 as shown in the last row and column of \autoref{tab:otherModelsOurData}.  Including post-processing, evaluation takes on average 1.8~seconds per page on a single core of an Apple M1 Max with 64 Gb of RAM.

\subsection{Benchmarks at high levels of localization (IOU=0.9)} \label{section:benchandgen}

As the ultimate goal of our method is the extraction of figures and their captions from scanned pages, we quantify how well our model performs on our dataset for a high degree of localization with an IOU cut-off of 0.9.

%\vspace{-10mm}
\autoref{tab:otherModelsOurData} shows how other deep learning models fair on our final test dataset.
Here, we use \textsf{ScanBank} \cite{scanbankthesis,scanbank} (based on \textsf{DeepFigures} \cite{deepfigures} and trained on a corpus of pre-digital electronic thesis and dissertations (ETDs)) and a version of \textsf{detectron2} \cite{wu2019detectron2} trained on the PubLayNet dataset \cite{publaynet}. \rc{\textsf{ScanBank} and \textsf{detectron2} are used for comparison as they are applied to raster-formatted articles (as opposed to vector-based methods like \textsf{pdffigures2} \cite{pdffigures2} which, applied to our data, results in F1 scores of $<$15\% in feature and final test splits). }
Both \textsf{ScanBank} and \textsf{detectron2} do not share our definitions of figures and figure captions exactly, thus to facilitate comparison we make some approximations and assumptions.

As discussed in \cite{scanbankthesis}, \textsf{ScanBank}'s figure's are defined as encompassing the figure caption, while our figure definitions exclude the caption.  In order to compare with our results, we initially performed metric calculations by re-defining our true figure boxes as the combination of figure and figure caption boxes when figure captions are associated with a figure.  However, we found that if we instead use our definitions of figures and captions, metrics from detections made with \textsf{ScanBank} are optimized, thus we use our definitions of figures and captions for all comparisons with this model. 
As \textsf{detectron2} does not find caption boxes specifically, but rather localizes generic ``text" boxes, we define \textsf{detectron2}-detected figure captions as those boxes with centers which are closest to a found figure's center.
To test the effects of our post-processing methods alone, we apply a subset of our post-processing steps to the results generated from both \textsf{ScanBank} and \textsf{detectron2} which we show for comparison to our method with and without post-processing in \autoref{tab:otherModelsOurData}. When applying post-processing to these other models' results, we use only up to the ``Step 5" described in \autoref{section:postprocessing} as we found this optimized the metrics for \textsf{ScanBank} and \textsf{detectron2} reported in \autoref{tab:otherModelsOurData}.

As shown in \autoref{tab:otherModelsOurData}, \textsf{ScanBank} does not perform well on our final test dataset with or without post-processing.
In particular, \textsf{ScanBank} does not detect captions reliably as the false negative rate (FN) is high.  
Additionally, there is a large portion of both figures and figure captions which are either erroneously detected, or not well localized as shown by the high false positive (FP) rate.
This is somewhat expected as the the ETD format is visually distinct from the articles in our dataset, including different fonts and caption placements.
When post-processing is applied the metrics for figure captions improve significantly with an increase of $\approx$30\% in true positive (TP) rate and decrease of $\approx$20\% in FP.

For the \textsf{detectron2} model without post-processing, TP rates are slightly higher than \textsf{ScanBank}'s,  
however FP rates remain comparable to \textsf{ScanBank}'s. 
FN rates are lower than our model (both with and without post-processing) by a few percentage points, likely due in part to the known differences in error profiles between YOLO-based (ours) and Mask-RCNN-based (\textsf{detectron2}) object detection models \cite{yolov1}.
Post-processing makes a large improvement on the TP rate for captions, increasing it by $\approx$35\% and decreasing FP by $\approx$40\%.  There is a modest increase in TP of $\approx$10\% for figures as well when post-processing is applied.

Post-processing (using all steps) has the largest effect on our model's results -- increasing TP rates of figures and captions by $\approx$25\% and $\approx$60\%, respectively.
This is not surprising as our post-processing method was developed using our scanned page training data.  Additionally, we employ a YOLO-based model which is used for detecting bounding boxes, not a segmentation method that would tend to produce larger TP rates at higher IOU thresholds -- the post-processing pipeline ``mimics" segmentation by changing box sizes to closer fit precise locations of caption words and figures, increasing overlap IOU. 

Taken together, the results of \autoref{tab:otherModelsOurData} suggest that other models generalize to our dataset at best moderately at high IOU, and only with application of our post-processing pipeline.
Because our post-processing steps require not only grayscale scanned pages, but their OCR outputs, this additional overhead (of both producing the OCR and post-processing steps) greatly reduces the gains in page processing speeds achieved with these other methods.

% THIS TABLE IS DONE
% comparison to other works
\begin{table}%[!htbp]
\begin{center}
\begin{tabular}{|c?cc?cc?cc?cc?cc?cc|}
    \hline
    & \multicolumn{2}{c?}{ScanBank}  & \multicolumn{2}{c?}{ScanBank}
    & \multicolumn{2}{c?}{detectron2$^\star$} & \multicolumn{2}{c?}{detectron2$^\star$} 
    & \multicolumn{2}{c?}{Ours} & \multicolumn{2}{c|}{Ours} \\
    & \multicolumn{2}{c?}{No PP} 
    & \multicolumn{2}{c?}{w/PP} 
    & \multicolumn{2}{c?}{No PP} & \multicolumn{2}{c?}{w/PP} 
    & \multicolumn{2}{c?}{No PP} & \multicolumn{2}{c|}{w/PP} \\
    & fig%$^*$ 
    & cap & fig & cap & fig & cap$^\dagger$ & fig & cap$^\dagger$ & fig & cap & fig & cap \\
    
    \hline
    TP &  
    69.9\unskip & 29.0\unskip &
    69.3\unskip & 52.8\unskip &
    72.0\unskip & 46.4\unskip &
    81.0\unskip & 80.9\unskip &
    58.2\unskip & 23.2\unskip &
    85.7\unskip & 86.7\unskip \\
    
    FP & 71.4\unskip & 28.8\unskip &
    43.6\unskip & 8.7\unskip &
    41.8\unskip & 68.2\unskip &
    27.1\unskip & 22.4\unskip &
    45.3\unskip & 82.3\unskip &
    13.7\unskip & 8.6\unskip \\

    FN &  1.7\unskip & 42.8\unskip &
    2.5\unskip & 40.7\unskip &
    0.6\unskip & 1.6\unskip &
    1.2\unskip & 4.9\unskip &
    3.1\unskip & 5.1\unskip &
    3.5\unskip & 6.0\unskip \\

    \thickhline 
    Prec &  49.5\unskip & 50.2\unskip &
    61.4\unskip & 85.9\unskip &
    63.3\unskip & 40.5\unskip &
    74.9\unskip & 78.3\unskip &
    56.2\unskip & 22.0\unskip &
    86.2\unskip & 90.9\unskip \\

    Rec &  97.6\unskip & 40.4\unskip &
    96.5\unskip & 56.5\unskip &
    99.2\unskip & 96.6\unskip &
    98.5\unskip & 94.3\unskip &
    95.0\unskip & 81.9\unskip &
    96.1\unskip & 93.6\unskip \\
    
    F1 &  65.7\unskip & 44.8\unskip &
    75.0\unskip & 68.1\unskip &
    77.2\unskip & 57.1\unskip &
    85.1\unskip & 85.6\unskip &
    70.6\unskip & 34.7\unskip &
    90.9\unskip & 92.2\unskip \\

    \hline
\end{tabular}
\end{center}
\vspace{-2mm}
\footnotesize{$^\star$ The tested version of \textsf{detectron2} is trained on the PubLayNet dataset \cite{wu2019detectron2}.}\\
\footnotesize{$^\dagger$ Here, captions are the ``text" classified box closest to the center of a figure.}\\
\vspace{-1mm} 
\caption{Performance metrics for %the deep learning detection methods 
\textsf{ScanBank} \cite{scanbankthesis,scanbank} 
and \textsf{detectron2} \cite{wu2019detectron2} for our final test dataset. IOU is 0.9. TP, FP, FN are in percentages of total true instances. Models with post-processing (``w/PP") and those without (``No PP") are shown for comparison. No retraining or transfer learning of \textsf{ScanBank} or \textsf{detectron2} have been done with our dataset. Errors from a 5-fold cross validation on all metrics are $\sim$1-2\%.}
\label{tab:otherModelsOurData}
\end{table}

%\subsection{\rc{FROM ORIG PAPER: Generalizability}}

\rc{This lack of generalizability is a known problem in the field of document layout analysis \cite{surveydeeplearning} and our model is no exception.  For example, when our model is applied to a selection of PubLayNet's non-commercial article pages \cite{publaynet}}\footnote{Non-commercial articles are necessary to access the high resolution scans and perform the OCR needed for our method.} \rc{we find F1 scores lower than those produced by \textsf{detectron2} for figures -- 65.7\% (ours) vs. 85.8\% (\textsf{detectron2}) for 207 figures.  Using the definitions of captions from \autoref{tab:otherModelsOurData} for 201 captions the results are more promising with F1 scores of 65.6\% (ours) to 65.5\%(\textsf{detectron2}).  Results for our (\textsf{ScanBank}'s) model applied to the \textsf{ScanBank} collection of ``gold-standard" ETDs \cite{scanbankthesis,scanbank} are lower overall with F1 scores of 25.5\%(38.4\%) for 197 figures and 16.3\%(1.4\%) for 140 captions.}
\rc{While these results suggest that our model may be more generalizable than other models for figure captions, tests on larger datasets are necessary for a firmer conclusion.}

% trying to standarsize spacing
\vspace{-1mm}
\section{Discussion and Future Work} \label{section:future}
%\vspace{-1mm}
This paper has focused on the localization of figures and figure captions in astronomical scientific literature.
We present results of a YOLO-based deep learning model trained on features extracted from the scanned page, \textsf{hOCR} outputs of the \textsf{Tesseract} OCR engine \cite{tesseract}, and the text processing outputs from spaCy \cite{spacy2}.  

Through ablation experiments we find the combination of the page and \textsf{hOCR} properties of (grayscale,  ascenders, decenders, word confidences, fraction of numbers in a word, fraction of letters in a word, punctuation, word rotation and spaCy POS) maximize our model's performance.
When compared to other deep learning models popular for document layout analysis (\textsf{ScanBank} \cite{scanbank,scanbankthesis} and \textsf{detectron2} \cite{wu2019detectron2}) we find our model performs better on our dataset, particularly at the high IOU thresholds (IOU=0.9) and especially for figure captions.
In particular, in line with our extraction goals, our model has relatively low false positive rates, minimizing the extraction of erroneous page objects.  \rc{For figures, our model does not perform well on other datasets (e.g. PubLayNet\cite{publaynet}, \textsf{ScanBank}\cite{scanbank}), however is as accurate or more accurate than others for figure captions.} 

Our work relies on a relatively small set of scanned pages ($\sim$6000).  While the results here for figures and captions surpass the estimates of $\sim$2000 instances per class required for training YOLO-based models \cite{bochkovskiy2020yolov4,Wang_2021_CVPR} our data contains many edge cases of complex layouts and we expect more data to improve results for these pages.  As our model relies on more than three feature channels, transfer learning on pre-trained YOLO-based models is less straight forward, but nonetheless could be a way to make use of our small dataset in future work.

Additionally, our current methodology does not make use of popular image processing features (e.g. connected components \cite{younas2019}) or loss functions/processing techniques that are ``non-standard" for YOLO-based methods \cite{non_nms}.  Future testing with the inclusion of these features may increase our model performance.

While all of our models converge within 150 training epochs, this is without the inclusion of any data augmentation.  
As our model uses not only grayscale but \textsf{hOCR} properties, typical data augmentation procedures (e.g. flipping, changes in saturation) are not appropriate for all feature layers.  However, it is likely that correctly-applied data augmentation will increase our model's performance. % from those metrics 
\rc{Our work would further benefit from a future large hyperparameter tuning study beyond the several values of learning rate tested in this paper.}

Finally, given the difference in error profiles between the YOLO-based method presented here and other Mask-RCNN/Faster-RCNN based \cite{yolov1} document layout analysis models (e.g. \textsf{detectron2}), it is likely that an ensemble model using both methods would further increase model performance.

This work is supported by a Fiddler Fellowship and a NASA Astrophysics Data Analysis Program Grant (20-ADAP20-0225). \rc{The authors thank the referees for their insightful input on this paper.}

%\rt{!!!!!!LIMIT IS 12 PAGES!!!!!}

\clearpage
\bibliographystyle{splncs04}
\bibliography{references}
\end{document}